\documentstyle[12pt,epsf]{article}
\def\half{\frac{1}{2}}
\def\be{\begin{equation}}
\def\ba{\begin{eqnarray}}
\def\ea{\end{eqnarray}}
\def\gr{\nabla}
\def\ee{\end{equation}}
\def\to{\rightarrow}
\def\tmu{$TH\epsilon\mu\;$}
\def\pd{\partial}
\def\nn{\not\!}
\def\sp{\overline\psi}

\def\al{\alpha}

\begin{document}

\begin{titlepage}
%\vspace*{4cm}
\begin{center}
{\Large\bf  The Equivalence Principle in the Non-baryonic
Regime}\vspace{1cm}\\
C. Alvarez
\footnotemark\footnotetext{E-mail:
calvarez@avatar.uwaterloo.ca} and 
R.B. Mann\footnotemark\footnotetext{E-mail:
mann@avatar.uwaterloo.ca}\\
Department of Physics \\
University of Waterloo\\
Waterloo, ONT N2L 3G1, Canada\\
\vspace{2cm}
\today\\
\end{center}
\vspace{2cm}
\begin{abstract}

We consider the empirical validity of the equivalence principle 
for non-baryonic matter.  Working in the context of the
\tmu formalism, we evaluate the constraints experiments place on
parameters associated with violation of the equivalence principle
(EVPs) over as wide a sector of the standard model as possible. 
Specific examples include new parameter constraints which
arise from torsion balance experiments, gravitational red shift,
variation of the fine structure constant, time-dilation measurements,
and matter/antimatter experiments.  We find several new bounds on
EVPs in the leptonic and kaon sectors.
\end{abstract}
\end{titlepage}

\section{Introduction and Summary}

The postulate that the equivalence between uniform acceleration and
a uniform gravitational field apply to all physical phenomena 
allowed Einstein to construct a theory of gravitation, 
general relativity, which revolutionized our conceptual understanding 
of the universe. It allowed a description of physics in which the effects
of gravitation are manifest as the dynamics of the geometry
of a curved spacetime. That this geometry is unique for all forms
of mass-energy is a consequence of Einstein's equivalence postulate.

Only decades later was it realized that this postulate is the foundation
for a rather broad class of theories of gravitation (which includes
general relativity) known as metric theories.  Any theory
of gravity that describes spacetime via a symmetric, second-rank 
tensor field $g_{\mu\nu}$ that couples universally to all 
non-gravitational fields respects the aforementioned equivalence 
between uniform acceleration and uniform gravitational fields, and
is by definition a member of this class. By definition,
non-metric theories break gravitational universality by
coupling additional gravitational fields to matter.

This understanding eventually resulted in a more precise formulation
of Einstein's equivalence postulate in terms of a number of physically
distinct principles \cite{Dicke}. The most basic of these is the
the Weak Equivalence Principle, or WEP, which states that the
trajectory of any test body (with a given initial velocity and
spacetime position) in a given gravitational field is independent of its 
internal structure or composition.  A natural extension of this
to include all nongravitational
phenomena states that, in addition to WEP, the outcomes 
of nongravitational test experiments
performed within a local, freely falling frame are independent 
of the frame's location 
(local position invariance, LPI) and velocity (local Lorentz 
invariance, LLI) in a background gravitational field. The combination
of WEP, LLI and LPI embody what is now known as the Einstein 
Equivalence Principle, or EEP.  The further extension of this
principle to include self-gravitating systems is known as the
Strong Equivalence Principle, or SEP.

Since a direct consequence of the EEP is that the outcome of local 
non-gravitational experiments should be independent of the effects of 
an external (slowly varying) gravitational field, direct tests 
of EEP may be carried out as follows. Consider an  Earth-based 
laboratory in which local nongravitational experiments are performed.  
External  gravitational potentials generated by the Earth, the
Sun, the planets, the Galaxy, {\it etc.} pervade this laboratory,
and any nonmetric couplings of these potentials to matter
can cause the outcomes of experiments to depend on the laboratory's  
position, orientation or velocity relative to these sources. 
This is a direct violation of (respectively) LPI and LLI.
The character of a violation reflects the form of the specific  nonmetric
coupling responsible for it.  It is only when LPI and LLI
are valid that local nongravitational  dynamics is indistinguishable
from special relativistic dynamics as  predicted by metric theories of
gravity.  

Tests of the validity of the various facets of the Equivalence
Principle have been carried out to impressive levels of precision.
The universality of free-fall (or UFF, a necessary consequence of
WEP) has been empirically validated to within $10^{-12}$.
Limits on violations of LLI to a precision of $\sim 10^{-22}$
have been imposed using laser experiments \cite{PLC}.
Recent experiments comparing the rates between H-maser and Hg clocks
have imposed a stringent new limit on LPI violation via
the bound $\dot{\alpha}/\alpha \leq 3.7\times 10^{-14}$ where
$\alpha$ is the fine-structure constant \cite{PTM}. Alternate
tests of LPI via gravitational redshift experiments 
could reach a precision up to $10^{-9}$ \cite{will1}.

Why, then, ought one to resist the temptation to conclude that
future experiments should ignore non-metric theories and
focus only on winnowing out the correct metric theory of gravity? 
There are four basic reasons.  One is the anticipated
improvements in precision of upcoming experiments by
as much as six orders of magnitude \cite{step}. If such
experiments yield improved limits on EEP-violation, this
will afford us a much greater degree of confidence in our physical
theories under the extreme conditions present
in many astrophysical and cosmological situations. Another is
historical: attempts to unify gravity with 
the other forces of nature have yielded a number of logically
possible, physically well-motivated, alternatives to general 
relativity which do not naturally respect the EEP \cite{Damour}.
A third reason is that tests of the EEP can provide us with
a unique way (perhaps the only way) of testing modern physical theories
that unify gravity with the other forces of nature insofar as such theories 
typically generate new interactions which violate
the equivalence principle \cite{string}.  Finally, EEP experiments to
date have probed effects that are predominantly sensitive to
nuclear electrostatic energy. Although violations of WEP/EEP due to
other forms of energy (virtually all of which are associated with baryonic
matter) have also been estimated \cite{HW}, the bulk of our empirical
knowledge about the validity of the equivalence principle is in the
baryon/photon sector of the standard model.

Comparatively little is known about the empirical validity of the EEP for
systems dominated by other forms of mass-energy
\cite{Hughes}.  Such systems include photons of differing
polarization \cite{Jody}, antimatter systems \cite{anti}, 
neutrinos \cite{utpal}, mesons \cite{Kenyon}, massive leptons \cite{grg}, 
hypothesized dark matter \cite{dark}, second and third generation matter, 
and quantum vacuum energies \cite{catshmpl}. There is no logically
necessary reason why such systems should respect any or all of WEP, EEP
or SEP. 

In order to establish the universal behavior of gravity,
we are therefore compelled to consider the validity of the EEP over 
as diverse a  range of non-gravitational interactions as is possible.
To this end, we consider in this paper an appraisal of the current
experimental basis for the equivalence principle in the context of the
\tmu formalism \cite{tmu}.  This formalism encompasses a broad class of 
non-metric theories of gravity and deals with the dynamics of charged 
particles and electromagnetic fields  in a static, 
spherically symmetric gravitational field. Non-metric couplings are
parameterized by a set of static spherically symmetric (SSS)
functions which take
on constant values for any physical systems whose spatio-temporal
extent is such that the gravitational field(s) may be considered
constant. The \tmu formalism thereby provides an
interpretative framework for EEP experiments, in that the results
of any given EEP experiment set bounds on (or alternatively
fix) these constant parameter values, henceforth referred to
as EEP-violating parameters, or EVPs. Low-energy effective field 
theories that arise from more fundamental theories of quantum gravity, 
unification and spacetime structure will typically yield specific 
predictions for such EVPs. By providing bounds on
these parameters, EEP tests give us an invaluable probe of the
fundamental physical laws that govern our universe.

Our approach differs from the more traditional usage of the \tmu 
formal\-ism,
which assumes a universal gravitational coupling for massive particles
that is distinct from the gravitational/electromagnetic coupling.
The only EVPs are the limiting speed of all massive bodies $c_0$
and the speed of electromagnetic radiation, $c_*$. For a theory
respecting LLI and LPI, $c_*=c_0$, and so in this framework EEP
experiments set bounds $|1-c^2_*/c^2_0| \equiv \xi$. Since we wish
to consider the possibility of EEP-violation for as many forms of
mass-energy as possible, we generalize this approach by breaking
universality for all species of particles. 
In this case the modified nongravitational
action reads
\ba\label{act1}
 S_{NG}&=&-\sum_a m_a\int dt\, (T_a-H_av_a^2)^{1/2}+\sum e_a \int dt\, v_a^\mu
 A_\mu(x_a^\nu)\nonumber\\
& &+ \half\int d^4x\,(\epsilon E^2- B^2/\mu),
\ea
where $m_a$, $e_a$, and $x_a^\mu(t)$ are the rest mass, charge, and
world line of particle $a$, $x^0\equiv t$, $v_a^\mu\equiv dx_a^\mu/dt$,
$\vec E\equiv-\vec\gr A_0-\pd\vec A/\pd t$,
$\vec B \equiv \vec\gr\times\vec A$. Violation of gravitational
universality in this formalism implies
that each given species of particle has its own distinct
gravitational coupling ({\it i.e.} its own metric), described 
by $g_{\mu\nu}^{(a)} = diag(T_a(r),-H_a(r),
-H_a(r), -H_a(r))$. These functions, along with $\epsilon$ and $\mu$
(which parameterize the metric for the electromagnetic field)
are arbitrary functions of the static spherically symmetric (SSS) 
(background) Newtonian gravitational potential
$U= GM/r$, which approaches unity as $U\to 0$. Expanding these
functions about the origin $X_0$ of the local frame of reference, it
is straightforward to see that, for systems in which spatio-temporal
variations of the \tmu functions can be neglected,
the limiting speed of the $a$-th species of massive particle  is
$c_a= (T_a(X_0)/H_a(X_0))^{1/2}$, and that the speed of electromagnetic
radiation ({\it i.e.} the speed of light) is $c_* = 
1/\sqrt{\epsilon(X_0)\mu(X_0)}$.  

Previous applications of the \tmu formalism assumed that all $c_a$'s 
were equal to $c_0$. In general this is not the case: the role of
experiment in this context is to provide quantitative information
on these EVPs or, more properly, on the ratios $c_a/c_*$, since the
action can always be rescaled by an overall constant. Since we expect
the symmetries implied by EEP to be at least approximate symmetries,
it is in practice more useful to consider bounds on the quantities
$\xi_a \equiv  |1-c^2_*/c^2_a|$. 

We therefore consider in this paper the bounds experiment places on
the various EVPs which follow from the action (\ref{act1}).

In the next section we review the present empirical evidence in
support of EEP and find constrains on violations of EEP among the different
species of massive particles. The main emphasis is to sort out
the empirical knowledge related to the validity of EEP in the 
non-baryonic sector of the standard model. We start reviewing the 
empirical limits on WEP violations as found by torsion balance 
experiments. We then  follow with
violations of LPI in the context of  gravitational red shift  experiments
and  the variation of the atomic fine structure constant. 
Next we review the time-dilation experiment (Hughes-Drever type) as 
performed by Prestage {\it et al.} \cite {PLC1},
in order to look for spatial anisotropy or LLI violations. 
We also consider limits on EVPs obtained from matter/antimatter 
experiments (CPT tests) as described by Hughes \cite{hughes1}. 
In the last section we
summarize and discuss the implications of our results.

\section {Empirical Review of the EEP}
\subsection{ E\"otvos type Experiments}

These experiments search for quantitative differences between the 
passive gravitational mass and the inertial mass of a given body.
The former is a dynamical quantity that determines 
the gravitational force acting on a body ({\it i.e.} its weight),
whereas the latter is a kinematical quantity that determines the response
of a body to any applied force. There is no logically necessary reason
why these quantities must be equal (in appropriate units), and so we
therefore expect
\be
m_p=m_I+\sum_A\eta^A E^A/c^2
\ee
where $E^A$ is the internal energy generated by interaction $A$, 
and $\eta^A$ is a dimensionless parameter that measures the strength 
of the WEP violation for body $A$. 

For two different bodies we can write the acceleration as
\be
a_1=(1+\sum_A\eta^A E_1^A/c^2)g\qquad a_2=(1+\sum_A\eta^A E_2^A/c^2)g
\quad .
\ee
A measurement on the relative difference in acceleration 
yields the so called ``E\"otvos ratio" given by
\be\label{etag}
\eta\equiv 2|\frac{a_1-a_2}{a_1+a_2}|=\sum_A\eta^A \left(\frac{E_1^A}{m_1c^2}
-\frac{E_2^A}{m_2c^2}\right)
\ee
Experiments carried out to date \cite{will1} imply that this ratio is 
constrained to be
\be\label{etal}
|\eta|<\left\{\begin{array}{r} 10^{-11}\\10^{-12}\end{array}\right.
\ee
This limit in turns constrains the violating parameter $\eta^A$ related to
each A-type interaction. This is possible provided the various
interactions do not conspire towards special types of cancellations so
that independent bounds can be gathered in
each case (see ref.\cite{will1} for quotations of those limits when referred
to interactions stemming from the atomic nucleus: strong, electrostatic,
magnetostatic, hyperfine,etc.).

This result can be analyzed in more detail if the inertial mass of an 
atom of atomic number $Z$ and mass number $A$ is written as \cite{Hughes}
\be
m_I(A,Z)=Z(m_e+m_p)+(A-Z)m_n+ E(A,Z)/c^2
\ee
and its gravitational mass as
\ba
m_G(A,Z)&=&(1+\delta_e)Zm_e+(1+\delta_p)Zm_p+(1+\delta_n)(A-Z)m_n
\nonumber\\&+&(1+\delta_E)E(A,Z)/c^2
\ea
where $E(A,Z)$ is the sum of all the binding energies 
and $\delta$ parameterizes possible
violations of WEP by each of the constituents. Then the following
limits are obtained \cite{Hughes}
\ba\label{etan}
|\delta_n|&<&5\times 10^{-9}\\\label{etae}
|\delta_e|&<&4\times 10^{-6}\\
|\delta_E|&<&5\times 10^{-9}
\ea
where the possibility of fortuitous cancelation is ignored. 
The limit on  $\delta_E$ constrains
the total atomic and nuclear binding energy. Note  that the independant
limit for electron quoted before is obtained  
provided $\delta_n=\delta_p$. Neutral atoms involved in
these experiments contain
equal number of protons and electrons, and so a more rigorous analysis
actually yields \cite{Hughes}: 
\be
|\delta_e\frac{m_e}{m_p}+\delta_p-\delta_n|< 2\times 10^{-10}
\ee
Since there is strong  evidence supporting the validity of EEP in
the baryonic sector and our aim is to extract as much information as
possible to the less explored leptonic domain, we omit consideration
of any gravitational anomaly stemming from baryonic matter.

In order to connect these results with the \tmu formalism, 
we study first the
case of a single particle falling in a SSS gravitational field ($U$). 
Variation of the first term of the action (\ref{act1}) gives the particle's 
equations of motion \cite{will1}:
\be
\frac{d\vec v_a}{dt}=\frac{d}{dt}\left(\frac{H_a}{W_a}v_a\right)+\half\frac{1}{W_a}
\gr(T_a-H_a v_a^2)
\ee
where  $W_a\equiv(T_a-H_a v_a^2)^{1/2}$. If we expand
the \tmu functions
\be\label{tu}
T(U)=T_0+T_0'\vec g_0\cdot\vec X+O(\vec g_0\cdot\vec X)^2
\ee
about the origin $X_0=0$ of the system, then
\ba
\frac{d\vec v}{dt}&=&-\half\frac{T_0'}{H_0}\vec g_0+\half\frac{H_0'}{H_0}g_0 v^2
\left(\frac{T_0'}{T_0}-\frac{H_0'}{H_0}\right)(\vec g_0\cdot\vec v)\vec v+\cdots
\nonumber\\&\equiv&\vec g_a
\ea
where  the species-labeling index-$a$ is implicit on  $T$, $H$ and $v$.

To lowest order we obtain for
the electron and proton the accelerations
\ba
\vec g_e&=&-\half\frac{T_e'}{H_e}\vec g_0+\cdots\\
\vec g_p&=&-\half\frac{T_p'}{H_p}\vec g_0+\cdots\label{gp}
\ea

The $\delta_a$ parameters introduced before quantify differences
in the acceleration of a given species of particle (electron, proton or
neutron in the present case) with respect  
to a standard $g$ (given by the choice
of units), that is, $\delta_a=|g_a-g|/g$.

The strongest constraint is on the neutron parameter (\ref{etan}), 
which is assumed to be
equivalent to that of the proton. This suggests a choice of units
in which $g\equiv g_p$, to the order given  in (\ref{gp}). In that case we
can write (c.f. (\ref{etae})):
\be\label{tmude}
\delta_e=|1-f(T)\frac{c_e^2}{c_p^2}|<4\times 10^{-6}
\ee
where $f(T)\equiv T_e'T_p/ T_eT_p'$, and $c_e$
and $c_p$ the respective limiting speeds. Note that within the 
traditional usage of the \tmu formalism 
there is no EEP violation at this level, and so the constraint
(\ref{tmude}) is trivially satisfied.

Turning next to the binding energy $E_A$, its dominant contribution
arises from the atomic nucleus. In the case of electromagnetic interactions
we can distinguish different internal energy contributions. For the
electrostatic  and magnetostatic nuclear energy, the violating EEP
parameters are bounded by \cite{will1} $\delta_{ES}<4\times 10^{-10}$
and $\delta_{MS}<6\times 10^{-6}$ respectively. A consideration
of these nuclear binding energies within the \tmu formalism
constrains the LPI violating parameters as \cite{will1}:
 \be\label{lim1}
|\Gamma_B|<2\times 10^{-10}\qquad |\Lambda_B|<3\times 10^{-6}
\ee
where
\be\label{para}
\Gamma_B\equiv\frac{T_B}{T_B'}\ln[\frac{T_B\epsilon^2}{H_B}]'\vert_{\vec X_0}
\qquad
\Lambda_B\equiv\frac{T_B}{T_B'}\ln[\frac{T_B\mu^2}{H_B}]'\vert_ {\vec X_0},
\ee
where the sub-index $B$ labels the metric related to baryonic matter ( which is
assumed to be the same among  baryons). Note that the former parameters 
are both equal to zero if the EEP is valid. The above result sets
 an upper limit on violations of WEP due to
a different gravitational coupling between baryonic matter and
electromagnetism.

The  electron played no role in the derivation of (\ref{lim1}).
In order to compare (gravitationally) electrons and photons we
should look at atomic binding energies. The electrostatic
interaction amongst the electrons themselves and between them
and the protons in the nucleus
is the dominant  form of energy in this case.
In general  these energies are within the range
of $E^{atom}\sim 10 eV$. Although the
experimental limit (\ref{etal}) is related via (\ref{etag}) to the
specific interaction under consideration, thereby involving
the difference between the atomic electrostatic energies
of the two bodies being tested (aluminum and platinum
in the case of the strongest limit quoted in (\ref{etal})), we can 
make a crude estimate by assuming
that this difference contributes as $E^{atom}/m_p\sim 10^{-8}$. This 
allows us to constrain the EVP related to
this interaction (the corresponding $\eta^A$), as $\delta_{atom}<10^{-3}$.
In the case of hydrogen atoms the \tmu formalism
implies  $\delta_{atom}=2\Gamma_e$, which is given by (\ref{para})
with the appropriate electron label. This result comes
after solving the Schr\"odinger equation (within the \tmu context)
for the principal atomic energy levels (it can be inferred from eq.(2.113) in ref.\cite{will1}).
{}From here we can extract  the approximate limit  $|\Gamma_e|<10^{-3}$ 
for the EVP associated with 
the relative gravitational coupling between electrons and 
electromagnetism. This limit is well below the analogous constraint for 
baryonic matter (\ref{lim1}).

\subsection{Gravitational Red Shift Experiments} 

In a redshift experiment the local energies at
emission $w_{em}$ and at reception $w_{rec}$ of a photon transmitted between
observers at different points in an external gravitational field are
compared in terms of
\be
Z=\frac{w_{em}-w_{rec}}{w_{em}} \equiv \Delta U\Big(1-\Xi\Big)
\ee
The anomalous redshift parameter ($\Xi$) measures the degree of 
LPI violation. It signals the breakdown of  the universality  of gravity,
and so depends on the nature of the transition involved in the experiment
(e.g., fine, hyperfine,etc..).

The most accurate test for the gravitational red shift corresponds
to the gravity probe A experiment \cite{redshift}, which was able to constrain
$|\Xi^{Hf}|<2\times 10^{-4}$. This experiment
employed hydrogen maser clocks, where the governing energy transition is given by
the hyperfine splitting due to  the interaction between the magnetic
moment (spin) of the nucleus (proton) and electron.

In the following we proceed to review this experiment assuming different 
gravitational couplings between electrons and protons 
(or baryons in general). In solving the Hydrogen atom, the 
relevant metric is the electron metric, the proton 
(or more generally, the nucleus in Hydrogenic atoms)
playing only the role of a static charge at rest with magnetic moment
$\vec \mu_p$.

The electromagnetic field produced by that source is
\cite{will1}
\be
A_0=-\frac{e}{\epsilon_0 r}\qquad\vec A=\half \mu_0\vec \mu_p\times\frac
{\vec r}{r^3}.
\ee
where the magnetic moment of the proton  is given by (in a manner
analogous to that for the electron):
\be
\vec \mu_p=\frac{T_B^{1/2}}{H_B}g_p\left(\frac{e\vec\sigma}{2 m_p}\right)
\ee
where the various parameters have their usual meaning.
Note that the proton metric affects only
the hyperfine splitting (due to $\vec \mu_p$),
since it arises from the  interaction between the magnetic
moments of the electron and proton (nucleus). 
The principal and fine structure atomic energy levels
depend only upon the electron metric. It is simple to check from 
ref.\cite{will1} that the hyperfine splitting scales as
\be\label{hf}
\Delta E_{hf}={\cal E}_{hf} \frac{{T_B}^{1/2}}{H_B}\frac{H_e^2}{T_e}
\frac{\mu_0}{\epsilon_0^3}
\ee
where ${\cal E}_{hf}$ depends on atomic parameters only.

If we expand (\ref{hf}) according to (\ref{tu}), then we can 
identify \cite{will1}
\be\label{hf1}
\Delta E_{hf}={\cal E}_{hf}(1-U)+{\cal E}_{hf} U\Xi^{hf}
\ee
with
\be
\Xi^{hf}=3\Gamma_B-\Lambda_B+\Delta
\ee
where we chose units such that the gravitational
potential is given by $U=-\half\frac{T_B'}{H_B}\vec g_0\cdot\vec X$,
and
 $\Gamma_B$ and $\Lambda_B$ are given by
(\ref{para}).
In (\ref{hf1}) we rescaled the atomic parameters to absorb the
\tmu functions and chose units such that $c_B=1$. The quantity $\Delta$ is
given by
\be
\Delta=2\frac{T_B}{T_B'}\left[2(\frac{H_B'}{H_B}-\frac{H_e'}{H_e})
-\frac{T_B'}{T_B}+\frac{T_e'}{T_e}\right]
\ee
and would vanish under the assumption that the leptonic and baryonic
\tmu parameters were the same.

The gravity probe A experiment constrains the corresponding LPI violating
parameter related to hyperfine transitions:
\be\label{xihf}
|\Xi^{Hf}|=|3\Gamma_B-\Lambda_B +\Delta|<2\;\times\;10^{-4}
\label{hyplpi}
\ee

Note that if  protons and electrons couple identically  to gravity then 
$\Delta=0$. On the other hand we can use the E\"otovos result
(\ref{lim1}) to assign the above limit to the $\Delta$ function only
and so have just a constraint on the relative baryonic and leptonic metrics.

\subsection{Variations of the Fine Structure Constant}

Experiments searching for a temporal variation of the
fine structure constant $\al$ can be divided into two categories:
cosmological and laboratory measurements. The first ones look for variations
within cosmological time scales and the others are based on clock 
comparisons over time durations of months or years. 

Laboratory measurements make use of clocks with ultrastable 
oscillators of differing physical composition, such as the superconducting 
cavity oscillator vs. cesium hyperfine clock transition. 
They rely on the ultrahigh stability of
the atomic standard clocks and set limits a few orders of magnitude less
stringent than the cosmological measurements. One of the most sensitive 
tests for $\al$--variation comes from the clock comparison 
between Hg$^+$ and H hyperfine transitions \cite{PTM}. This 
experiment set an upper bound of $\dot\al/\al\leq 3.7\times 
10^{-14}/yr$ after a 140 day observation period.
 Note that any variation of $\al$, whether a
cosmological time variation or a spatial variation via a dependence 
of $\al$ on the gravitational potential, will force a variation in 
the relative clock rates between any such pair of clocks. 

 The result of  ref. \cite{PTM}
is based on the increasing importance of relativistic contributions
to the hyperfine splitting as Z increases in the group I alkali elements and
alkali ions. Using the theoretical expressions for the
hyperfine splitting in hydrogen ($a_h$) and the alkali
atom  or ion ($A_{Hg^+}$ in this case), along with the
experimental result for the drift between the two clocks, namely
$d\ln(A_{Hg^+}/a_h)/dt< 2.1(0.8)\times 10^{-16}/day$, they 
constrained:
\be\label{al1}
\frac{1}{\al}\frac{d\al}{dU}\frac{dU}{dt}\leq 3.7\times 10^{-14}/yr
\ee
where we have explicitly shown the position dependence of $\al$
via the gravitational potential $U$. Note that within the
\tmu formalism $\al$ rescales according to:
$$\al\to\frac{\al}{\epsilon_0}\sqrt{\frac{H_e}{T_e}}$$
where $T_e$ and $H_e$ are the electron metric locally evaluated
at the center mass position of the atom (for example $\vec X=0$).
If these functions along with $\epsilon$ are expanded according to
(\ref{tu}), then
\be
\al\to\al(1+\Gamma_e U)
\ee
where $\Gamma_e$ is defined by (\ref{para}), and 
adequate units are used to define the gravitational potential.
It is clear that we can evaluate:
\be
\frac{1}{\al}\frac{d\al}{dU}=\Gamma_e+O(U)
\ee

A crude estimation for the time variation of the gravitational potential, 
which can be considered as the solar gravitational potential at the 
laboratory, is obtained by taking the  extreme variation of $U$ over the 
140 day period, that is the seasonal change, 
$\Delta U\sim 3\times 10^{-10}$ \cite{Hughes}. This altogether bounds:
\be\label{gael}
|\Gamma_e|<5\times 10^{-5}
\ee

Baryonic matter contributes to the ratio $A_{Hg^+}/a_h$  only via
the  nuclear $g$ factors related to each atom. These parameters
are determined mainly by the strength of the strong interactions.
These interactions were assumed to obey EEP, when  deriving eq. (\ref{al1})
from the empirical variation of the hyperfine ratio.
Hence the purely leptonic contribution to the change on $\al$
allows us to improve the upper bound for the LPI violating parameter
related to electrons (relative to photons), as stated in eq. (\ref{gael}).

\subsection{Time Dilation Experiments}

LLI requires that the local, nongravitational physics of a bound system of
particles be independent of its velocity and orientation relative to any preferred frame.
If LLI were violated the energy levels of a bound system such as a nucleus
could be shifted in a way that correlates the motion of the bound particles
in each state with  the preferred direction, leading  to an 
orientation--dependent  binding energy (anisotropy of inertial mass).  The most precise
experiments of this sort  \cite {PLC} search for a time dependent quadrupole
splitting of Zeeman levels. They compare the nuclear-spin- precession
frequencies between two gases with nuclear spin $I=3/2$ and $I=1/2$,
 the latter being insensitive to a quadrupole splitting. These results place
the constraint 
$$
\xi_B=(1-c_*^2/c_B^2)<\,6\,\times \,10^{-21}
$$
on the relative gravitational coupling between 
electromagnetism and baryonic matter. This result stems from 
nuclear interactions only and so the leptonic metric plays
no role in its derivation.

A weaker bound was obtained by Prestage {\it al.} \cite{PLC1} by comparing 
the frequency of a nuclear spin-flip transition in $^9Be^+$ to the
frequency of a hydrogen maser transition. In the context of the
standard \tmu formalism the EEP violating anomalies coming from
the hyperfine transitions of both clocks are negligible in comparison
to those originating from the electric quadrupole moment of 
the  $^9Be^+$  nucleus. The derived limit of $\xi_B<10^{-18}$ 
was obtained under the assumption that
the relative gravitational interaction between 
electromagnetism and electrons is the same as that between
electromagnetism and baryons.

In the following we drop this assumption and
proceed to interpret this experiment accordingly. Experiments which
search for the time dependent quadrupole Zeeman splitting suggest
that  $\xi_B \approx 0$. We shall assume this for simplicity, so that
we have $c_e\not=c_B \approx c_*$. In other words, we
analyze the effects from EEP violations in the electron/photon sector
of the standard model.

Let us review the result of ref.\cite{PLC1}. They studied the ratio of the
frequency of clocks defined by  the hyperfine transition of the $^9Be^+$ 
(nuclear spin-flip) and $H$ (electron spin-flip) atoms. They searched for
variation in the clock frequencies of the form:
\be
\nu=\nu_0+ A_2P_2(\cos\Theta)
\ee
where $P_2(\cos\Theta)=3(\cos^2\Theta-1)/2$, and $\Theta$ is the angle 
between  the quantization axis and the direction of matter anisotropy
in the nearby universe.

The limit  $A_2< 10^{-5}$ Hz was obtained
from looking for variations with respect to
motion through the mean rest frame of the universe. 
On the other hand the \tmu formalism predicts a change of the form:
\be\label{nub}
\delta\nu=\nu-\nu_0=\xi_B\delta\nu^{nuclear}_{ES} P_2(\cos\Theta)
\ee
where $\delta\nu^{nuclear}_{ES}$ corresponds to the nuclear electrostatic energy coming
from the electric quadrupole moment of the $^9Be^+$ nucleus. This is reoriented by
the clock transition which flips the nuclear spin and therefore contributes
to the energy transition. Comparing the former expression 
with the
experimental result for $A_2$ yields the previously quoted limit
on $\xi_B$.
Note that the nonmetric anomalies stemming from the hyperfine
transition of the $^9Be^+$ and maser clocks were ignored in 
deriving (\ref{nub}).
By including them and neglecting the baryonic contributions, we obtain
an expression of the form
\be\label{nul}
\delta\nu=\xi_e\delta\nu^{atom}_{Hf} P_2(\cos\Theta)
\ee
where $\xi_e=1-c_B/c_e$ and $\delta\nu^{atom}_{Hf}$ accounts for the 
hyperfine anomalies mentioned above. These can be expressed as
\be\label{nul1}
\delta\nu^{atom}_{Hf}=(A_{Be}-A_{H})\nu_0
\ee
where the $A$ -coefficients account for LLI violations stemming from
the hyperfine structure related to each atom (similar to the time
dilation coefficients introduced by  Gabriel and Haugan \cite{gabriel}),
and $\nu_0$ for the clock transition being measured.

In the \tmu framework we can find an expression for $A$, which to lowest
order gives: $A_{Be}-A_{H}=-V^2/2$ (see appendix for details),
 where $V$ is the Earth velocity with respect to the preferred frame.
Assuming $\xi_B=0$,  the Prestage experiment implies
the limit
\be
\xi_e(A_{Be}-A_{H})\nu_0\sim A_2<10^{-5} \hbox{[Hz]}
\ee
when looking for variations with respect to
motion through the mean rest frame of the universe ($V\sim 10^{-3}$).
This, along with the $^9Be^+$ clock frequency value $\nu_0=303$ MHz,
imposes the constraint \be
\xi_e< \frac{2}{3}\times 10^{-7}
\ee
on the leptonic EEP violating parameter.

\subsection{Antimatter/matter experiments}
The universality of  gravity  embodied by EEP makes no distinction between  particles
and antiparticles. This assumption goes beyond the CPT theorem, because the
gravitational acceleration of a particle is not an intrinsic property but involves
its interaction with the external gravitational field.

 Hughes made use of experiments testing CPT-violation in the Kaon
system to deduce an improved test of the WEP for the 
antiproton \cite{hughes1}, via potentially differing gravitational 
couplings of quarks and antiquarks. We wish here to review the influence 
of a possibly anomalous gravitational interaction on neutral kaons within 
the context of the \tmu formalism.

In the presence of an external gravitational field the non gravitational action for the
kaon system can be expressed as \cite{hughes1}
\be\label{actk}
S=-\int^t dt'\frac{{\bf M}}{\gamma}
\ee
with
\be 
\gamma^{-1}=\sqrt{g_a^{\mu\nu}\frac{dx_\mu}{dt}\frac{dx_\nu}{dt}}
\ee
where the gravitational interaction of particle $a$ 
($K_0$ or $\bar K_0$) is mediated by
\be
g_a^{\mu\nu}=diag(T_a, -H_a, -H_a, -H_a),
\ee
and $x_\mu$ are the space-time coordinates along the kaon's world line. ${\bf M}$ is the mass
matrix , whose diagonal terms correspond to the $K_0$ and $\bar K_0$ masses. \footnotemark
\footnotetext{Actually
eq. (\ref{actk}) is a short notation  to express the dependence of 
the kaon and anti-kaon  on the gravitational field, since $\gamma$ 
is also dependent on particle $a$, and so  factors
each diagonal term of ${\bf M}$ in a different way. The off-diagonal
terms of ${\bf M}$ are irrelevants since the experimental data to be used 
later on  relates the diagonal elements alone.}

We assume that the gravitational field changes slowly along the dimensions 
of the laboratory system, so that we can expand the $TH$ functions 
according to (\ref{tu}). In that case
\be\label{expg}
g^{\mu\nu}=\eta_*^{\mu\nu}+U h_*^{\mu\nu}+O(U^2)
\ee
with
\be
\eta_*^{\mu\nu}=diag(T_0, -H_0, -H_0, -H_0)\qquad
 h_*^{\mu\nu}=diag(T_0', -H_0', -H_0', -H_0'),
\ee
and $U=\vec  g_0\cdot\vec X$ as the gravitational potential yielding 
the gravitational acceleration $\vec g_0$. The $a$-label related to 
each particle is implicit on each $TH$ parameter. 
The former expansion reduces to  Hughes' work provided  $T_0=H_0=1$ and 
$T_0'= -H_0'=2\alpha$, where $\alpha$ is dependent 
on the species of particle.

By using (\ref{expg}) we can approximate
\be
\gamma^{-1}=\gamma^{-1}_*+\frac{U}{2}\gamma_*t_*+O(U^2)
\ee
where
\be
\gamma^{-1}_*=\sqrt{\eta_*^{\mu\nu}\frac{dx_\mu}{dt}\frac{dx_\nu}{dt}}
\qquad
t_*=h_*^{\mu\nu}\frac{dx_\mu}{dt}\frac{dx_\nu}{dt},
\ee
which in turns reduces the Lagrangian to (see footnote)
\be\label{lag}
L=-{\bf M}\left[\gamma_*^{-1}+\frac{U}{2}\gamma_*t_*+O(U^2)\right]
\ee

Hughes argued that for the experiment under consideration the potential
$U$ can be regarded as constant, and so the gravitational terms can be absorbed
into redefinitions of the mass matrix elements. The apparent potential and
velocity dependence of those elements can be partially removed by introducing
physical (or local) time and length units (which are also influenced by gravity), 
that are gathered from instruments that are assumed to obey EEP. 
Actually this is irrelevant when considering the difference between 
the kaon and antikaon masses, since if the instruments 
obey EEP they are going to affect  the kaon and
anti-kaon system in the same way and therefore it will give no 
contribution to the mass difference. 

In our case, in order to make the redefinition of masses possible, we need to introduce
the standard of units. We have chosen
units such that the speed of light is equal to one, 
and so a kaon system obeying the EEP can be reduced to the form:
\be\label{lag1}
L=-\frac{{\bf M}}{\gamma_0}
\ee
with $\gamma_0=(1-v^2)^{-1/2}$. Note that  $\gamma_*=T_0^{-1/2}
(1-v^2/c_a^2)^{-1/2}$, where $c_a=(T_0/H_0)^{1/2}$ corresponds to the
local limiting speed of the massive $a$--type particle. We can then
expand
\be
\gamma_*=T_0^{-1/2}\gamma_0(1-\frac{\xi_a}{2}v^2\gamma_0^2)+O(\xi_a^2)
\ee
with $\xi_a=(c_a^2-1)/c_a^2$.

Hence,  by comparing (\ref{lag}) to (\ref{lag1}),we can introduce an effective mass, up to
$O(\xi_a)O(U)$ of the form:
\be\label{mass}
m_a^{(eff)}=m_a T_0^{1/2}\left(1+\frac{\xi_a}{2}v^2\gamma_0^2+\frac{U}{2}\gamma_0^2\frac{t_*}{T_0}
\cdots\right)
\ee
The CERN experiment with kaons at energies $\sim 100$ GeV constrained (see \cite{hughes1} and references
therein) :
\be
|(m(K^0)-m(\bar K^0))/m(K^0)|<5\times 10^{-18}
\ee
which along with (\ref{mass}) bounds the difference between the limiting speed of the kaon and the
anti-kaon as:
\be\label{limk1}
|\xi_K-\xi_{\bar K}|\simeq|c_K^2-c_{\bar K}^2|<1.5\times 10^{-22}
\ee

Moreover if we assume that there is no fortuitous cancelation between the two EEP violating
terms in (\ref{mass}), we can constrain the $O(U)$ contribution as:
\be
\frac{U}{2}\gamma_0^2\left|\frac{t_*}{T_0}-\frac{\bar t_*}{\bar T_0}\right|<5\times 10^{-18}
\ee
By taking the supercluster potential at the surface of the Earth 
to be $U\sim 3\times 10^{-5}$,
we can bound the relative non-universal gravity coupling in the neutral kaon system as:
\be\label{limk2}
|\frac{T'_0}{T_0}-\frac{\bar T'_0}{\bar T_0}+\frac{\bar H'_0}{\bar T_0}-\frac{ H'_0}{T_0}|
<10^{-17}
\ee

Note that eqs. (\ref{limk1}) and (\ref{limk2}) bound the relative metric 
between the neutral kaon and antikaon in terms of the $TH$ functions and 
their derivatives, evaluated locally.  In Hughes' analysis there is no 
EEP violation at the level  described by eq. (\ref{limk1}).

In the absence of gravity the $TH$ functions approach unity, restoring
CPT  symmetry. We note that violation of EEP in the kaon-antikaon
sector need not imply violation of CPT, as has recently been demonstrated
\cite{Ham}.

\section{Concluding Remarks}

Up to now, the strongest constraints on the EEP have involved
the baryonic sector of the standard model relative to the 
electromagnetic one. This is not surprising given the relative strength 
of $10^6$ between nuclear and atomic interactions. We have
demonstrated in this paper that many of the experiments which
test EEP actually provide us with a broader degree of empirical
information over the different sectors of the standard model. However
the constraints on EEP-violation in these other sectors are typically 
much weaker than those in the baryon sector. 
This difference is manifest in the limits imposed on 
LPI-violating parameters related to baryons and electrons,
which are extracted from torsion balance experiments. These tests 
look for anomalous gravitational accelerations for an entire atom, 
and so the effects of possible EEP violations
coming from  electron-nucleus interactions will be suppressed by the
total atomic mass. 

In this respect, gravitational redshift experiments provide cleaner
tests for leptonic matter insofar as its relative gravitational coupling 
to electromagnetism is concerned. Such experiments
probe atomic energy transitions, where 
the nucleus plays the role of a static
electromagnetic source. In the case of  experiments involving the 
principal, fine or Lamb shift transitions, the effect of any anomalous 
redshift will be to constrain only the relative gravitational
coupling between electrons and photons \cite{catlamb}. 
In those energy transitions the effect of the
nuclear magnetic moment (and so the nuclear metric) can be neglected. 
However the most precise gravitational redshift experiment 
employed hydrogen maser clocks, where the energy shift is
produced by the magnetic interaction between the nucleus and the electron, 
and so a hybrid bound is obtained (\ref{xihf}). 

By reviewing  laboratory tests for variations of the fine structure 
constant, we were able to obtain the strongest bound on the leptonic LPI 
violating parameter. Although it involved
hyperfine transitions, the empirical output was the hyperfine ratio between 
two atoms and so cancelled  out effects of the nuclear metric 
(assumed to be the same among nucleons).

The experiment of Prestage ({\it et al.}) has the distinct feature amongst
Hughes-Drever type experiments of employing hyperfine transitions 
stemming from the electron-nucleus interaction, as opposed to
nucleon-nucleon ones. A re-analysis of the results of this experiment
yielded a bound on the leptonic LLI violating parameter to a level
comparable to those coming from Lamb shift or $g-2$ experiments \cite{grg}.

Finally, we employed the \tmu formalism within the context 
of matter/antimatter experiments. We set new constraints on EEP violations 
coming from CPT tests in the neutral kaon-antikaon system. 
The remarkable limit of $10^{-22}$ was obtained for
the difference between the limiting speed of kaons/antikaons.

The empirical validity of the EEP
must be checked separately for each sector of
the standard model. We have shown in this paper how existing experiments
can provide us with some empirical information to this end in
several non-baryonic sectors.  However it is the proposal and design of 
new experiments  that probe the validity of the EEP in the 
non-baryonic sector that will ultimately provide us with the best 
empirical foundation for a metric description of gravity. It is our
hope that this paper will further motivate such work.

\section*{Acknowledgments}

This work was supported by the Natural Sciences and Engineering
Research Council of Canada.

\section* {Appendix} 

The spacetime scale of atomic systems allows one to ignore the spatial variations
of $T$, $H$, $\epsilon$, $\mu$, and evaluate them at the center of mass
position of the system, $\vec X=0$.  After rescaling coordinates, charges,
and electromagnetic potentials, the field theoretic extension of the
action (\ref{act1}) can be written in the form
\begin{equation}\label{act}
S=\int d^4x\, \sp(i\nn Q-m-\xi_*\gamma^0Q_0)\psi
 + \half\int d^4x\,(E^2-B^2),
\end{equation}
where local natural units are used, $Q_\mu\equiv=p_\mu-eA_\mu$, 
$\nn\! Q=\gamma_\mu Q^\mu$, and $\xi_*=(1-c_*/c_e)$ ,
with  $c_*$ and $c_e$ as the local speed of light and
limiting speed of electrons respectively. Note that  local natural
units were used in action (\ref{act}), by taking $\hbar=1$, and $c_*=1$ (which is
taken to be identical to the baryonic limiting speed $c_B$).

It is clear that under a Lorentz transformation, the electromagnetic part
of (\ref{act}) is invariant but not the fermion--photon 
interaction sector, which
adds to the total density Lagrangian a non metric term of the form:
\be\label{inter1}
{\cal L}_\xi=-\xi_*\gamma^2\sp\beta\cdot\gamma\beta\cdot Q\psi
\ee
where $\gamma^2\equiv 1/(1-\vec V^2)$ and  $\beta^\mu\equiv(1,\vec V)$;
henceforth $\beta^2\equiv 1-\vec V^2$, with $V$  the relative velocity 
between the
preferred frame defined by (\ref{act}) 
and the laboratory system (Earth).

We are interested in the non relativistic terms of 
(\ref{inter1}). These are
\be\label{inter2}
{\cal L}_\xi\simeq-\xi_*\gamma^2\left[\varphi^\dagger\frac{\vec\sigma}{2m}\cdot(\vec\gr
\times(\beta\cdot Q)\vec\beta)\varphi+\cdots\right]
\ee
where $\varphi$ represents the large component of the Dirac spinor and
the ellipsis denotes terms that do not depend on the 
electron spin.

{}From (\ref{inter2}) we can  read the interaction term between the
 electron spin and an external magnetic field, which added to the 
metric contributions gives
up to $O(\xi_*)O(V^2)$:
\be\label{inter3}
H_{hf}=-\vec\mu_e\cdot\left[\vec\gr\times\vec A+
\xi_*\vec\gr\times(\vec V\cdot \vec A)\vec V\right]
\ee
where in units of $\mu_B=e/2m_e$, $\vec\mu_e=g_e\mu_B\vec J$,
$\vec J$ being the electron spin ($J=1/2$). Recall that the magnetic field 
produced by the magnetic moment $\vec\mu_N=g_N\mu_B\vec I$ of the nucleus 
is $\vec A=-\frac{1}{4\pi}\vec\mu_N\times\vec\gr\frac{1}{r}$, 
where $\vec I$ represents the nuclear spin ($I=1/2$ for hydrogen). 
We have also introduced the
corresponding gyromagtenic ratios for  the electron and nucleus.

The $^9Be^+$ atom can be treated as an alkali atom, where the main
interaction governing the hyperfine transitions comes from the magnetic
moment interaction between the valence electron and the nucleus only. Hence
it can be considered as a hydrogenic atom with $I=3/2$ and $J=1/2$. The
effects of the electrons in the closed shell are accounted for in the
values for the principal quantum number and effective charge of the atom
which are empirically determined \cite{atomic}.

Let us start reviewing the Maser clock. Here the relevant atomic transition
takes place under a weak magnetic field, which does not break the
coupling between the electron and nuclear spin. That is the atomic states
can be labeled by the total angular momentum $\vec F=\vec J+\vec I$ and
the corresponding quantum number $M_F$. The hyperfine transition in this
case is described by $\Delta F=1$, $\Delta M_F=0$. 

It is straightforward within this \tmu formalism 
to obtain from (\ref{inter3}) the corresponding
hyperfine transition
\be
\nu_{hf}^{(H)}=\nu_0\left[1+\xi_*\frac{V^2}{6}\left(1+P_2(\cos\Theta)\right)
\right]
\ee
where $\nu_0=-\langle\vec\mu_e\cdot\vec\gr\times\vec A\rangle|_{\Delta F=1,
\Delta M_F=0}$.
Since experiments search for frequency changes of the form
\be\label{paraA}
\delta\nu=\xi_* A\nu_0 P_2(\cos\Theta)
\ee
the parameter $A$ in (\ref{nul1}),
for the maser clock is identified as $A_{H}=\frac{V^2}{6}$.

The situation is different for the $^9Be^+$ clock. Here the relevant atomic
transition occurs in the presence of a strong magnetic field 
that breaks the nuclear electron spin coupling. The main energy 
contribution to the electron comes from the electron spin interaction 
with that external field and not with the field of the nucleus. 
Hence the atomic states are described by the quantum numbers 
$M_I$ and $M_J$. 
The $^9Be^+$ hyperfine transition
corresponds to a nuclear resonance with $\Delta M_J=0$, $\Delta M_I=1$.

After computing  (\ref{inter3}) for the relevant states we obtain 
\be
\nu_{hf}^{(Be)}=\nu_0\left[1+\xi_*\frac{V^2}{6}\left(4-2P_2(\cos\Theta)\right)
\right]
\ee
for the modified hyperfine 
transition for the  $^9Be^+$ clock, 
where $\nu_0=-\langle\vec\mu_e\cdot\vec\gr\times\vec A\rangle|_{\Delta M_J=0,
\Delta M_I=1}$. We neglected the nuclear spin interaction with the 
external magnetic field, since it contributes as $O(m_e/m_p)$ 
when $B\sim 1[T]$. 
By identifying the former result with (\ref{paraA})
 we finally obtain: $A_{Be}=-\frac{V^2}{3}$.

\end{document}